\documentclass[aps,prb,twocolumn,showpacs]{revtex4}

\usepackage{amsmath}
\usepackage{color}
\usepackage{amssymb}

\usepackage{graphicx}

\begin{document}

\title{Reentrant superconductivity in URhGe}

\author{V.P.Mineev}
\affiliation{Commissariat a l'Energie Atomique, INAC / SPSMS, 38054 Grenoble, France}

\begin{abstract}

There is presented a phenomenological description of phase diagram  of ferromagnet superconductor URhGe.
In frame of   the Landau phenomenological theory it was found that phase transition between anisotropic ferromagnetic and paramagnetic states under strong enough magnetic field perpendicular to direction of easy magnetization 
 changes    from the second to the first order type.  It is shown that magnetic susceptibility corresponding to longitudinal magnetic fluctuations strongly increases  in vicinity of the first order transition stimulating reentrance of superconducting state.
    The reentrant superconductivity observed near the first order transition line at temperatures about twice lower than the tricritical point temperature exists both in ferromagnet and paramagnet state. The critical temperature of transition to superconducting state falls down at intersection with line of ferromagnet-paramagnet phase transition.  

\end{abstract}

\date{\today}
\maketitle

\section{Introduction}

The superconductivity in uranium compounds UGe$_2$, UCoGe and URhGe (for the most recent reviews, see Refs. 1 and 2) developing deeply inside of ferromagnetic state most probably belongs to
triplet equal spin pairing type. These strongly anisotropic ferromagnets have orthorhombic crystal structure with magnetic moment directed along one of crystallographic directions: parallel to $a$-axis in UGe$_2$ and to $c$-axis in UCoGe and URhGe.  The latter has  peculiar property.  At low enough temperature  the magnetic field about 1.3 Tesla 
 directed along $b$-axis suppresses the superconducting state\cite{Hardy} but at much higher field about 8 Tesla the superconductivity is recreated and exists 
 till the field about 13 Tesla.\cite{Levy} The  maximum of superconducting critical temperature in this field interval is $\approx 0.4~K$.

The problem of stimulation of triplet superconductivity in anisotropic ferromagnets by magnetic field perpendicular to spontaneous magnetization has been discussed by the present author \cite{Mineev2011,Mineev2014}.  This has allowed to explain the strong upturn of the upper critical field along $b$ direction  in UCoGe \cite{Aoki2009} by the increase of the pairing interaction caused by the suppression of the Curie temperature by magnetic field.
There was pointed that a similar mechanism  also stimulates the recreation  superconducting state in URhGe. This case, however, the superconducting state exists not only inside of ferromagnetic state but also in the paramagnetic state separated from the ferromagnetic state by the phase transition of the first order. 

The observation of abrupt collapse  of spontaneous magnetization under strong enough external field along $b$ axis has been reported already in the first publication about magnetic field-induced superconductivity in ferromagnet URhGe.\cite{Levy} Quite recently the first order character of transition has been directly confirmed by the observation of hysteresis \cite{AokiKnebel2014} in the Hall resistivity in vicinity of transition field $H_R\sim 12.5~T$ at temperatures below \\0.8 K.

In the present paper we develop the approach of Refs.5,6 in application to superconductivity arising on both sides of the first order type phase transition from anisotropic ferromagnetic to paramagnetic state  under external magnetic field perpendicular to spontaneous magnetization.  In  next section we present the Landau type phenomenological description of the ferromagnet-paramagnet phase transition. Then we   find the components of magnetic susceptibility determining the superconducting pairing interaction and show that magnetic susceptibility corresponding to longitudinal magnetic fluctuations strongly increases  in vicinity of the first order transition stimulating reentrance of superconducting state. 
After that the  difference in  the superconducting critical temperature  on both sides ferromagnet-paramagnet phase transition   is demonstrated . 
 The results are discussed in the last part of the paper.

\section{Phase transition in orthorhombic ferromagnet under  magnetic field perpendicular to spontaneous magnetization}

We begin with
the Landau free energy of orthorhombic ferromagnet 
in
magnetic field ${\bf H}({\bf r})$
\begin{equation}
{\cal F}=\int d V(F_M+F_\nabla),
\label{FE}
\end{equation}
where in 
\begin{eqnarray}
F_M=\alpha_{z}M_{z}¥^{2}+\beta_{z}¥M_{z}¥^{4}+\delta_zM_z^6~~~~~~~~~~~~~~~~~\nonumber\\
 +\alpha_{y}M_{y}^{2}+\alpha_{x}M_{x}¥^{2}+\beta_{yz}¥M_{z}¥^{2}¥M_{y}¥^{2}¥+\beta_{xz}¥M_{z}¥^{2}¥M_{x}¥^{2}-{\bf M}{\bf  H},
\label{F}
\end{eqnarray}
we bear in mind the orthorhombic anisotropy and also the term of the sixth order in powers of $M_z$.
The density of gradient energy is  taken in exchange approximation
\begin{equation}
F_\nabla=\gamma_{ij}\frac{\partial {\bf M}}{\partial x_i}\frac{\partial {\bf M}}{\partial x_j}.
\label{nabla}
\end{equation}
Here, the $x, y, z$ are directions of the spin axes pinned to $a, b, c$
crystallographic directions correspondingly, 
\begin{equation}
\alpha_{z}=\alpha_{z0}(T-T_{c0}), 
\end{equation}
$\alpha_x>0$, $\alpha_y>0$ and matrix $\gamma_{ij}$ 
is 
\begin{equation}
\gamma_{ij} = \left(\begin{array}{ccc} \gamma_{xx} & 0 & 0\\
0 & \gamma_{yy} & 0 \\
0 & 0 & \gamma_{zz}
\end{array} \right).
\end{equation}

In constant magnetic field ${\bf H}=H_y\hat y$
the equilibrium magnetization projections along $x,y$ directions are obtained by minimization of free energy (\ref{F}) in respect of $M_x,M_y$
\begin{equation}
M_x=0,~~~~M_y=\frac{H_y}{2(\alpha_y+\beta_{yz}M_z^2)}.
\label{My}
\end{equation}
 Substituting these expressions back to (\ref{F})
we obtain
\begin{eqnarray}
F_M=\alpha_{z}M_{z}¥^{2}
+\beta_{z}M_{z}^{4}+\delta_zM_z^6-\frac{1}{4}\frac{H_y^2}{\alpha_y+\beta_{yz}M_z^2},
\label{F1}
\end{eqnarray}
that gives after expansion of denominator in the last term 
\begin{equation}
F_M=-\frac{H_y^2}{4\alpha_y}+\tilde\alpha_{z}M_{z}¥^{2}
+\tilde\beta_{z}¥M_{z}¥^{4}+\tilde\delta_zM_z^6+\dots,
\label{F2}
\end{equation}
where
\begin{eqnarray}
&\tilde\alpha_{z}=\alpha_{z0}(T-T_{c0})+\frac{\beta_{yz}H_y^2}{4\alpha_y^2},\\
&\tilde\beta_{z}=\beta_z-\frac{\beta_{yz}}{\alpha_y}\frac{\beta_{yz}H_y^2}{4\alpha_y^2}\\
&\tilde\delta_{z}=\delta_z+\frac{\beta_{yz}^2}{\alpha_y^2}\frac{\beta_{yz}H_y^2}{4\alpha_y^2}
\end{eqnarray}

We see that under a magnetic field perpendicular to direction of spontaneous magnetization  the Curie temperature decreases as
\begin{equation}
T_c=T_c(H_y)=T_{c0}-\frac{\beta_{yz}H_y^2}{4\alpha_y^2\alpha_{z0}}.
\label{Cur}
\end{equation}
The coefficient $\tilde\beta_z$ also decreases with $H_y$ and reaches  zero at
\begin{equation}
H_y^{cr}=\frac{2\alpha_y^{3/2}\beta_z^{1/2}}{\beta_{yz}}.
\end{equation}
At this field under fulfillment the  condition
\begin{equation}
\frac{\alpha_{z0}\beta_{yz}T_{c0}}{\alpha_y\beta_z}>1
\end{equation}
the Curie temperature (\ref{Cur}) is still positive and at $$H_y>H_y^{cr}$$  phase transition from paramagnetic to ferromagnetic state becomes the transition of the first order. The point $(H_y^{cr},T_c(H_y^{cr}))$ on the line  paramagnet-ferromagnet phase transition is tricritical point.

The minimization of the free energy Eq. (\ref{F2}) gives the value of the order parameter in ferromagnetic state
\begin{equation}
M_z^2=\frac{1}{3\tilde\delta_z}[-\tilde\beta_z+\sqrt{\tilde\beta_z^2-\tilde\alpha_z\tilde\delta_z}].
\end{equation}
The minimization of the free energy in paramagnetic state
\begin{equation}
F_{para}=\alpha_yM_y^2-H_yM_y
\label{p}
\end{equation}
in respect $M_y$ gives equilibrium value of magnetization projection on axis $y$ in paramagnetic state
\begin{equation}
M_y=\frac{H_y}{2\alpha_y}.
\end{equation}
Substitution it back in Eq. (\ref{p}) yields the equilibrium value of free energy in paramagnetic state
\begin{equation}
F_{para}=-\frac{H_y^2}{4\alpha_y}.
\end{equation}
On the line of the phase transition of the first order
determined by  equations \cite{StPhys}
\begin{equation}
F_M=F_{para},~~~~~\frac{\partial F_M}{\partial M_z}=0
\end{equation}
the order parameter $M_z$ has the jump
\begin{equation}
 M_z^{\star^2}=-\frac{\tilde\beta_z}{2\tilde\delta_z}.
 \label{jump}
\end{equation}
Its substitution  back in equation $F_M=F_{para}$ gives the equation of the first order transition line
\begin{equation}
4\tilde\alpha_z\tilde\delta_z=\tilde\beta_z^2,
\label{line}
\end{equation}
that is
\begin{equation}
T^\star=T^\star(H_y)=T_{c0}-\frac{\beta_{yz}H_y^2}{4\alpha_y^2\alpha_{z0}}+\frac{\tilde\beta_z^2}{4\alpha_{z0}\tilde\delta_z}.
\end{equation}
The corresponding negative jump of $M_y$ is given by
\begin{equation}
M_y^\star=\frac{H_y}{2(\alpha_y+\beta_{yz}M_z^{\star^2})}-\frac{H_y}{2\alpha_y}.
\end{equation}

\section{susceptibilities}

Magnetic susceptibilities along all directions are found as this has been done in \cite{Mineev2011,Mineev2014}.
In paramagnetic state they are:
\begin{eqnarray}
\chi^p_{xx}({\bf k})\cong \frac{1}{2\alpha_x+\gamma_{ij}k_ik_j},\\
\chi^p_{yy}({\bf k})\cong \frac{1}{2\alpha_y+\gamma_{ij}k_ik_j},\\
\chi^p_{zz}({\bf k})\cong \frac{1}{2\tilde\alpha_z+\gamma_{ij}k_ik_j}.
\end{eqnarray}
The susceptibilities  in ferromagnetic state are
\begin{eqnarray}
&\chi_{xx}({\bf k})=\chi^p_{xx}({\bf k}),\\
&\chi_{yy}({\bf k})\cong \frac{1}{2(\alpha_y+\beta_{yz}M_z^2)+\gamma_{ij}k_ik_j},\\
&\chi_{zz}({\bf k})\cong \frac{1}{2\tilde\alpha_z+12\tilde\beta_zM_z^2+30\tilde\delta_zM_z^4+\gamma_{ij}k_ik_j}.
\end{eqnarray}
The spin-triplet pairing interaction is expressed through the odd part of the susceptibility components\cite{Mineev2011,Mineev2014} :
\begin{equation}
\chi^u_{ii}({\bf k},{\bf k}')=\frac{1}{2}[\chi_{ii}({\bf k}-{\bf k}')-\chi_{ii}({\bf k}+{\bf k}')],~~~~i=x,y,z.
\end{equation}
Thus we have for paramagnetic state
\begin{eqnarray}
 \chi^{pu}_{xx}({\bf k},{\bf k}^\prime)\cong \frac{\gamma_{ij}k_ik_j^\prime}{2(\alpha_x+\gamma k_F^2)^2},\\
\chi^{pu}_{yy}({\bf k},{\bf k}^\prime)\cong \frac{\gamma_{ij}k_ik_j^\prime}{2(\alpha_y+\gamma k_F^2)^2},\\
\chi^{pu}_{zz}({\bf k},{\bf k}^\prime)\cong \frac{\gamma_{ij}k_ik_j^\prime}{2(\tilde\alpha_z+\gamma k_F^2)^2}.
\end{eqnarray}
Here $\gamma k_F^2\approx\gamma_{ij}k_ik_j$.
The corresponding components in ferromagnetic state are:
\begin{eqnarray}
&\chi^{u}_{xx}({\bf k},{\bf k}^\prime)=\chi^{pu}_{xx}({\bf k},{\bf k}^\prime),\\
&\chi^{u}_{yy}({\bf k},{\bf k}^\prime)\cong \frac{\gamma_{ij}k_ik_j^\prime}{2(\alpha_y+\beta_{yz}M_z^2+\gamma k_F^2)^2},\\
&\chi^{u}_{zz}({\bf k},{\bf k}^\prime)\cong \frac{\gamma_{ij}k_ik_j^\prime}{2(\tilde\alpha_z+6\tilde\beta_zM_z^2+15\tilde\delta_zM_z^4+\gamma k_F^2)^2}.
\end{eqnarray}
 Thus, at transition from paramagnetic to ferromagnetic state the $y$ and $z$ components of susceptibility abruptly decrease their values. 

As it was demonstrated in \cite{Mineev2011,Mineev2014}  pairing interaction in ferromagnetic state  is mostly determined by 
$z$ component of susceptibility.   Making use Eqs.(\ref{jump}), (\ref{line}) one can find its value at line of the first order transition
\begin{equation}
\chi^{u}_{zz}({\bf k},{\bf k}^\prime)\cong \frac{\gamma_{ij}k_ik_j^\prime}{2(4\tilde\alpha_z(T^\star)+
\gamma k_F^2)^2}= \frac{\tilde\gamma_{ij}\hat k_i\hat k_j}{2(4\frac{T^\star-T_c(H_y)}{T_{c0}}+
1)^2}
\label{fer1}
\end{equation}
Here, 
 $\tilde\gamma_{xx}=\gamma_{xx}/\gamma,\dots$ are numbers of the order of unity, $\hat k_i=k_i/k_F$ are components of the unit vector ${\bf k}$ and,
 as in Ref.6, we have used the estimation $\frac{\alpha_{z0}}{\gamma k_F^2}\cong\frac{1}{T_{c0}}$.

 It is instructive    to compare  this with
$z$ component of susceptibility at $H_y=0$ found in Ref.6
\begin{equation}
\chi^{u}_{zz}({\bf k},{\bf k}^\prime)\cong \frac{\gamma_{ij}k_ik_j^\prime}{2(-2\alpha_z+
\gamma k_F^2)^2}= \frac{\tilde\gamma_{ij}\hat k_i\hat k_j^\prime}{2(2\frac{T_{c0}-T}{T_{c0}}+
1)^2}.
\label{fer2}
\end{equation}

The denominators in these two equations have quite different magnitude. Indeed, taking into account that superconductivity in URhGe exists at $T\ll T_{c0}$ one can note that 
\begin{equation}
2\frac{T_{c0}-T}{T_{c0}}\approx 2,
\end{equation}
at the same time  the magnetic fields at line of the first order are just slightly larger than the field in tricritical point $H^{cr}$. Hence
\begin{equation}
4\frac{T^\star-T_c(H_y)}{T_{c0}}=\frac{\tilde\beta_z^2}{\alpha_{z0}\tilde\delta_zT_{c0}} \ll 1.
\end{equation}

Thus the susceptibility in vicinity of the first order transition proves to be much larger than susceptibility at small transverse fields $H_y$.
This protects the reentrance of superconductivity near the first order transition.

\section{Superconducting state in vicinity of the first order transition}

In Ref.5,6 there was shown that the  suppression of the Curie temperature by magnetic field perpendicular to spontaneous magnetization leads to effective increase of pairing interaction. This effect can in principle compensate the suppression of superconductivity by the orbital depairing. In UCoGe where the Curie temperature $T_c$  not strongly exceeds the temperature of transition to superconducting state $T_{sc}$ this mechanism successfully explains  the upturn of the upper critical field or nonsensitivity  $T_{sc}(H)$ 
for field along $b$ direction above 
5 Tesla. 

In URhGe the Curie temperature is much higher than $T_{sc}$. Hence, the orbital effect  succeeds to suppress the superconducting state  ($H_{c2}^b(T=0)\approx 1.3~T$ see Ref. 3)
much before the effect of 
decreasing of Curie temperature and stimulation of pairing intensity reveals itself.  But at fields higher than 8 Tesla the latter effect starts to overcome the orbital depairing and the superconducting state recreates. The critical temperature of superconducting transition begins grow up and approaches to the line of the first order transition from ferromagnetic to paramagnetic state and intersects it.\cite{Levy,AokiKnebel2014} 
Here we have a look what is going on with line of superconducting phase transition at intersection with line of ferromagnet-paramagnet first order phase transition $T^\star(H_y)$.

If the external field  orientated along  b-axis, that is perpendicular to the exchange field $h$,
 it is natural to choose the spin quantization axis along the  direction of the total magnetic field $
 h\hat z+H_y\hat y$. Then the normal state  matrix Green function is  diagonal 
\begin{equation}
\hat G_n=\left( \begin{array}{cc}G^{{\uparrow}}& 0\\ 
0 & G^{\downarrow}
\end{array}\right ),
\end{equation}
where
\begin{equation}
G^{{\uparrow,\downarrow}}=\frac{1}{i\omega_n-\xi^{{\uparrow,\downarrow}}_{{\bf k}}\pm\mu_B\sqrt{h^2+H_y^2}
}.
\end{equation}

In view large spin-up spin-down band splitting one can neglect by the interband $\Delta^{\uparrow\downarrow}$ pairing amplitudes
and take the matrix of the order parameter  as
\begin{equation}
\hat \Delta=\left( \begin{array}{cc}\Delta^{{\uparrow}}& 0\\ 
0& \Delta^{\downarrow}
\end{array}\right )
\end{equation}

The critical temperature of transition in superconducting state $T_{sc}(H_y)$ is determined from self-consistency equations linear in the spin-up $ \Delta^\uparrow$ and the spin-down $\Delta^\downarrow$ order parameter components.
In ferromagnetic state they are
 \cite{Mineev2011,Mineev2014}:  
\begin{widetext}
\begin{eqnarray}
&\Delta^{\uparrow}({\bf k},{\bf q})
=\mu_BI^2T
\sum_{n}
\sum_{{\bf k}' }
\left\{\left[\chi^u_{zz}({\bf k},{\bf k}')\cos^2\varphi+\chi^u_{yy}({\bf k},{\bf k}')\sin^2\varphi\right]
G_{1}^\uparrow
G_{2}^\uparrow
\Delta^{\uparrow}({\bf k}',{\bf q})\right.\nonumber\\
&+\left.\left[
(\chi^u_{xx}({\bf k},{\bf k}')-\chi^u_{yy}({\bf k},{\bf k}'))\cos^2\varphi+(\chi^u_{xx}({\bf k},{\bf k}')-\chi^u_{zz}({\bf k},{\bf k}'))
\sin^2\varphi\right]
G_{1}^\downarrow
G_{2}^\downarrow
\Delta^{\downarrow}({\bf k}',{\bf q})
\right\},
\label{e11}
\end{eqnarray}
\begin{eqnarray}
&\Delta^{\downarrow}({\bf k},{\bf q})
=\mu_BI^2T
\sum_{n}
\sum_{{\bf k}' }
\left\{\left[(\chi^u_{xx}({\bf k},{\bf k}')-\chi^u_{yy}({\bf k},{\bf k}'))\cos^2\varphi+(\chi^u_{xx}({\bf k},{\bf k}')-\chi^u_{zz}({\bf k},{\bf k}'))\sin^2\varphi\right]
G_{1}^\uparrow
G_{2}^\uparrow
\Delta^{\uparrow}({\bf k}',{\bf q})\right.\nonumber\\
&+\left.\left[\chi^u_{zz}({\bf k},{\bf k}')\cos^2\varphi+\chi^u_{yy}({\bf k},{\bf k}')\sin^2\varphi\right]
G_{1}^\downarrow
G_{2}^\downarrow
\Delta^{\downarrow}({\bf k}',{\bf q})
\right\}.
\label{e21}
\end{eqnarray}
\end{widetext}
Here, $I\approx T_{c0}/nm^2$ is an exchange constant, $n$ is the concentration of uranium atoms, $m$ is magnetic moment per atom at zero temperature,
$G_{1}^\uparrow=G^{\uparrow}({\bf k}',\omega_n)$, 
$G_{2}^\uparrow=G^{\uparrow}(-{\bf k}'+{\bf q},-\omega_n)$ and similarly for the  $G_{1}^\downarrow$ and 
$G_{2}^\downarrow$ Green functions.
Angle $\varphi$
is determined by
$$\tan\varphi={H_y}/{h}.$$

In paramagnetic state $\varphi=\pi/2$, $\chi^u\to\chi^{pu}$,
\begin{equation}
G^{{\uparrow,\downarrow}}=\frac{1}{i\omega_n-\xi_{{\bf k}}\pm\mu_BH_y}
\end{equation}
and the system of equations is reduced to
\begin{widetext}
\begin{eqnarray}
&\Delta^{\uparrow}({\bf k},{\bf q})
=\mu_BI^2T
\sum_{n}
\sum_{{\bf k}' }
\left\{\chi^{pu}_{yy}({\bf k},{\bf k}')
G_{1}^\uparrow
G_{2}^\uparrow
\Delta^{\uparrow}({\bf k}')
+
(\chi^{pu}_{xx}({\bf k},{\bf k}')-\chi^{pu}_{zz}({\bf k},{\bf k}'))
G_{1}^\downarrow
G_{2}^\downarrow
\Delta^{\downarrow}({\bf k}',{\bf q})
\right\},
\label{e11}
\end{eqnarray}
\begin{eqnarray}
&\Delta^{\downarrow}({\bf k},{\bf q})
=\mu_BI^2T
\sum_{n}
\sum_{{\bf k}' }
\left\{(\chi^{pu}_{xx}({\bf k},{\bf k}')-\chi^{pu}_{zz}({\bf k},{\bf k}'))
G_{1}^\uparrow
G_{2}^\uparrow
\Delta^{\uparrow}({\bf k}',{\bf q})
+\chi^{pu}_{yy}({\bf k},{\bf k}')
G_{1}^\downarrow
G_{2}^\downarrow
\Delta^{\downarrow}({\bf k}',{\bf q})
\right\}.
\label{e21}
\end{eqnarray}
\end{widetext}
As we mentioned already the $y$ and $z$ components of susceptibilities undergo a finite jump at first order phase transition  from ferromagnetic to paramagnetic state.
The  Fermi surfaces of split spin-up and spin-down electron bands, and  the average density of states on them  also undergo an abrupt changes. Finally, the structure of equations for determination of
the critical temperature of transition in superconducting state $T_{sc}(H_y)$  is quite different on both  sides of the ferromagnet-paramagnet phase transition. 
So,  the line of $T_{sc}(H_y)$  
should  undergo a jump at intersection of line of the first order phase transition $T^\star(H_y)$. 
The amplitude of spin up-up pairing in ferromagnet state is mostly determined by large $z$ component of susceptibility but this is not the case in paramagnet state where it is determined by much smaller $y$ component. As result the critical temperature of superconducting transition abruptly fall down at transition from ferromagnet to paramagnet state.
The experiment \cite{Levy} clearly demonstrates an abrupt fall of the critical temperature at this transition.

\section{Concluding Remarks}

In application to 
URhGe  we have presented the phenomenological description of phase diagram in this orthorhombic ferromagnet  under magnetic field along $b$ axis perpendicular to direction of spontaneous magnetization.
Making use   the Landau phenomenological theory we have found that phase transition between anisotropic ferromagnetic and paramagnetic states under strong enough magnetic field perpendicular to direction of easy magnetization 
 changes  its order  from the second to the first order type. The tricritical point $(H_y^{cr}, T_c(H_y^{cr}))$ on phase transition  line divides it on two peaces  
 $T_c(H_y)$ and $T^\star(H_y)$ of the second and the first order correspondingly.
 
The appearance of reentrant superconductivity is explained by  strong increase of magnetic susceptibility corresponding to longitudinal magnetic fluctuations in vicinity of the first order transition.  
 The reentrant superconductivity observed in vicinity of first order transition line $T^\star(H_y)$ at temperatures about twice lower than $T_c(H_y^{cr})$ exists both in ferromagnet and paramagnet state. The critical temperature of transition to superconducting state undergoes an abrupt falldown at intersection with line of ferromagnet-paramagnet phase transition.  
 
  It should be  noted that similar treatment can be undertaken at magnetic field directed along $a$-axis. Howewer, $a$-direction is magnetically much harder
than $b$ one: $\alpha_x\gg\alpha_y$. Hence, the suppression of the Curie temperature  and $\tilde\beta_z$ coefficient by magnetic field $H_x$  is much less effective. However, one can expect  developement of similar phenomena  at much higher fields   along $x$ direction.
 By the same reason if we already are in vicinity of the phase transition of the first order caused by application of magnetic field along $b$-axis and then add some field along $a$ axis increasing the magnitude of total magnetic field $H=\sqrt{H_y^{\star^2}+H_x^2}$ this introduces negligible changes in pairing interaction realized in absence of field along $a$-axis. At the same time the upper critical field in $a$ direction is in one and half times larger than in $b$ direction.\cite{Hardy} This roughly explains the stability of reentrant superconductivity in URhGe 
up to the fields $H=\sqrt{H_y^{\star^2}+H_x^2}\approx 30$ Tesla.\cite{Levy2007}

It is known that   in presence of an external field along direction of spontaneous magnetization $\hat z$ the line of the first order transition $T^\star(H_y)$ spreads to two surfaces of the first order transition $T^\star(H_y,\pm H_z)$. At these surfaces the jump in ferromagnet spontaneous magnetization
decreases and disappears completely on some lines 
 beginning at tricritical point  $T_c(H_y^{cr},H_z=0)$. 
There was proposed\cite{Levy} that these lines
are finished at zero temperature in some quantum critical points in $(H_y, H_z)$ plane. 
 The critical magnetic fluctuations have been put forward as a source stimulating superconductivity in vicinity of line of ferromagnet-paramagnet first order phase transition $T^\star(H_y)$ in absence of $H_z$. This idea looks like plausible. Nevertheless one 
can remark that in general tricritical line $T^{cr}(H_y,H_z)$  can never reach the zero temperature  or simply be dislocated far enough from the superconducting region on the phase diagram. Here we have demonstrated that reentrant superconductivity in URhGe can arise even in absence of critical fluctuations due to drastic increase of longitudinal susceptibility in vicinity
 on the first order transition line from paramagnet to ferromagnet state.
 
  The treatment of superconducting properties of uranium ferromagnets developed in Ref. 5,6 and in the present paper has been performed  in frame of weak coupling superconductivity theory. We have operated with frequency independent magnetic susceptibilities found in  thermodynamically equilibrium state.  Certainly, there are many properties of these materials like field dependence of  electron effective masses\cite{Aoki12,Aoki14}, NMR and NQR relaxation anisotropy\cite{Hattory2012,Hattory2014}, temperature dependence of the upper critical fields \cite{Aoki2009} which demand development of theory of 
  frequency dependent electron - magnon interaction determining superconducting pairing  in magnetically strongly anisotropic media.

\end{document}